# Quantum critical behavior in magic-angle twisted bilayer graphene

Alexandre Jaoui[1,*], Ipsita Das[1], Giorgio Di Battista[1], Jaime Díez-Mérida[1], Xiaobo Lu[1], Kenji Watanabe[2], Takashi Taniguchi[2], Hiroaki Ishizuka[3,4], Leonid Levitov[4] and Dmitri K. Efetov[1,*]

1. ICFO - Institut de Ciencies Fotoniques, The Barcelona Institute of Science and Technology, Castelldefels, Barcelona, 08860, Spain.
2. National Institute for Material Sciences, 1-1 Namiki, Tsukuba, 305-0044, Japan.
3. Department of Physics, Tokyo Institute of Technology, Meguro, Tokyo, 152-8551, Japan.
4. Department of Physics, Massachusetts Institute of Technology, Cambridge MA 02139, USA.

*Correspondence to: alexandre.jaoui@icfo.eu and dmitri.efetov@icfo.eu.

**Abstract**: The flat bands [1] of magic-angle twisted bilayer graphene (MATBG) host strongly-correlated electronic phases such as correlated insulators [2-6], superconductors [7-11] and a strange metal state [12]. The latter state, believed to be key for understanding the electronic properties of MATBG, is obscured by various phase transitions and thus could not be unequivocally differentiated from a metal undergoing frequent electron-phonon collisions [13]. Here, we report transport measurements in superconducting MATBG in which the correlated insulator states are suppressed by screening. The uninterrupted metallic ground state shows resistivity that is linear in temperature over three decades and spans a broad range of doping including those where a correlation-driven Fermi surface reconstruction occurs. This strange-metal behavior is distinguished by Planckian scattering rates and a linear magneto-resistivity. In contrast, near charge neutrality or a fully-filled flat band, as well as for devices twisted away from the magic angle, we observe the archetypal Fermi liquid behavior. Our measurements demonstrate the existence of a quantum-critical phase whose fluctuations dominate the metallic ground state throughout a continuum of doping. Further, we observe a transition to the strange metal upon suppression of the superconducting order, suggesting a relationship between quantum fluctuations and superconductivity in MATBG.

Conventional metals, in which electron interactions are described by Landau's Fermi-liquid theory, feature a hallmark temperature-dependent resistivity $\rho \propto T^2$ and a $\rho \propto B^2$ magneto-resistivity [14]. However, a strikingly different behavior arises in the presence of strong electron correlations, where alongside superconductors, magnets and insulators, unusual 'strange' metal phases can survive down to $T \to 0$, as reported in cuprates [15,16], ruthenates [17], pnictides [18], heavy-fermion systems [19] and, recently, in twisted di-chalcogenides [20]. Such 'strange' metal phases display the unique dependences $\rho(T) \propto T$ and $\rho(B) \propto B$ and are associated with ultra-fast carrier scattering governed by the universal Planckian dissipation rate $1/\tau = k_B T/\hbar$, with $\hbar$ the reduced Planck's constant and $k_B$ the Boltzmann's constant [21]. The observation of $\rho(T) \propto T$ down to $T \to 0$ signals the proximity to a quantum critical point: the neighboring metallic ground state is dominated by inelastic scattering with electronic quantum-critical fluctuations. These fluctuations have been mostly related to magnetic ordering [22], yet purely nematic fluctuations were also reported [23]. Understanding the intricate relationship between quantum fluctuations, finite-temperature 'strange' metallicity and the exotic phase transitions found in a wide variety of strongly-correlated systems is a major conundrum in condensed matter physics.

Early measurements in twisted bilayer graphene rotated by the magic angle of $\theta = 1.1°$ (MATBG) observed a metallic phase at temperatures 0.5 K < T < 30 K with a linear $\rho \propto T$ [12,13] and Planckian scattering rates in close proximity to correlated-insulating [2-6] and superconducting phases [7-11] for electron filling factors of $\nu = \pm 2$ per moiré unit cell. First interpretations suggested a 'strange-metal' phase [12] emerging from purely electronic interaction [24] and drawing an analogy to other strongly correlated systems [15-23]. Subsequently, however, it was proposed that a conventional electron-acoustic phonon scattering mechanism [25] above a characteristic Bloch-Grüneisen temperature $T_{BG}$ (typically, exceeding a few kelvin), could also result in a similar behavior under favorable assumptions [13, 26]. An investigation of the $T \to 0$ regime of MATBG would allow to definitively differentiate between these distinct scenarios, yet, such studies have so far been impeded by the abundant low-temperature insulating and superconducting phase transitions.

This manuscript reports on comprehensive electronic transport measurements extending down to 40 mK, firmly establishing the existence of a 'strange' metal phase. In order to reveal the low-temperature metallic states, we deliberately chose MATBG devices with ultra-close metallic screening layers and twist-angles that slightly deviate from 1.1° [9, 27]. In particular, we focus on device D1 with a twist-angle $\theta = 1.04°$ and a screening layer spacing of $d = 9.5$nm (inset of Fig.1.a), which did not show correlated-insulating phases (in the hole-doped region) but still showed a robust superconducting dome (see Fig.1.a). We performed resistivity measurements for a broad parameter space of $(n, T, B)$ with carrier density $n$ tuned across the entire flat-band region, temperatures from $T = 40$ mK to $T = 20$ K and magnetic fields up to $B = 1$ T (see SI section B for extended measurements up to 100 K). Our measurements confirm the existence of a $T$-linear resistivity in the center of the flat-bands above $T > 1$ K, in agreement with previous results [12, 13], and firmly establish its uninterrupted continuation down to 40 mK, which cannot be explained by conventional electron-phonon collisions. Further, we unveil another signature of the 'strange' metal state: a linear $B$-dependence of the resistivity. In contrast, we find a typical Fermi-liquid behavior $\rho \propto (T^2, B^2)$ in the vicinity of the band edges, and in twisted bilayer graphene (TBG) with twist-angles that strongly deviate from the magic-angle $\theta > 1.3°$ (see SI section C for more details). Lastly, we demonstrate that the 'strange' metal state extends into the superconducting dome region after suppressing superconductivity by an applied magnetic field. These observations establish the existence of a quantum-critical continuum and demonstrate a strong resemblance between MATBG and a variety of quantum-critical systems, pointing to an intricate relationship between electronic quantum fluctuations and superconducting/correlated insulating states. While recent findings of the isospin Pomeranchuk effect in MATBG [28, 29] point to soft spin and valley fluctuations as a driving mechanism of the low-$T$ phase diagram, the nature of the quantum fluctuation in MATBG remains an open question.

Fig. 1.a shows the four-terminal resistivity $\rho$ of device D1 as a function of the moiré band filling factor $\nu = n/n_0$ with $n_0 = 1/A_0$ ($A_0$ is the area of the moiré unit cell) for temperatures ranging from 40 mK up to 20 K. We observe insulating behavior, i.e. an increasing resistivity when temperature is decreased, at electrostatic doping levels which correspond to the vicinity of the charge neutrality point $\nu = 0$, for the fully-filled flat band $\nu = \pm 4$ and around $\nu = +3$. Superconducting domes are found close to half-filling ($\nu = \pm 2$). By choosing a screening layer separation that is smaller than a typical Wannier orbital size of 15nm [9] we were able to quench the correlated insulators at $\nu = -2$ and leave the hole-doped region entirely metallic (apart from the superconducting dome). Clearly resolved is also the isospin Pomeranchuk effect, where the resistance peaks at $\nu = \pm 1$ are more pronounced at elevated $T$ [28, 29]. The simplicity of the phase diagram of the hole-doped D1 allows for an in-depth study of the metallic ground state and its ties to the neighboring superconducting ground state. Similar datasets measured on other devices (listed in section A of the SI) are shown in the sections B and C of the SI and draw a consistent picture of the metallic ground state.

We first discuss the temperature dependence of the resistivity and its evolution with the filling factor. A basic zero-order picture emerges from the (numerical) derivative $\left(\partial \rho / \partial T\right)_\nu$ shown in Fig. 1.b : this quantity is roughly temperature independent in a wide range of filling factors and is weakly sensitive to doping. The details of this behavior are illustrated in Fig.1.c which present the resistivity vs. temperature for successive filling factors. Starting from the insulating regime at the charge neutrality point, metallicity is recovered at $\nu \approx -0.15$, which first shows a super-linear temperature dependence below 15 K, and then saturates into a linear dependence. With increased doping, the onset of the linear dependence is quickly shifted to lower temperatures. Starting from $\nu \approx 2$, the $T$-linear regime extends down to the base temperature and remains $T$-linear until a second super-linear regime is found for $\nu < -3.5$. This 'strange'-metal phase is only interrupted by a superconducting transition observed around half-filling.

We analyze the temperature dependence below $T < 10$ K by fitting the resistivity with $\rho(T) = \rho_0 + A_{T,\gamma} T^\gamma$, where the parameters $A_{T,\gamma}$ and $\gamma$ define the prefactor and the exponent of the $T$-dependence while $\rho_0$ is the residual resistance at $T = 0$. Fig 1.d shows $(\rho(T) - \rho_0)$ on a log-log scale, which allows to trace $\rho(T)$ over three orders of magnitude down to centikelvin temperatures. We fit each curve with $A_{T,\gamma} T^\gamma$, which results in linear lines on the log-log plot, with a slope that is directly defined by $\gamma = \partial [ln(\rho(T) - \rho_0)] / \partial [ln(T)]$. In proximity of the charge neutrality point ($\nu \approx -0.2$) and of full filling ($\nu = -3.7$), we find a $\gamma = 2 \pm 0.1$ which results in a super-linear $\rho(T) \propto T^{2\pm0.1}$ dependence. However, for the filling factor range of $-3.5 < \nu < -2$ we find $\gamma = 1$, which gives rise to a strictly linear $\rho(T) \propto T$ dependence with a high slope of $A_{T,1} > 0.25$ kΩ/K, which is shown in the inset of Fig.1.d. Strikingly the $T$-linear dependence extends without interruption from the base temperature of 40 mK to a temperature of 10 K, above which it saturates (see the sections D and E of the SI for a discussion on this saturation upon reaching the Mott-Ioffe-Regel limit).

Can the observed linear dependence be explained by electron-phonon scattering? The electron-phonon mechanism yields a weak $T$-linear resistivity only above the Bloch-Grüneisen temperature with typical values $T_{BG} > 10K$, whereas below $T_{BG}$ the $T$-dependence is superlinear, $\rho_{e-ph} \propto T^4$ [26]. The temperature of 40mK at which the $T$-linearity is observed in device D1 is several orders of magnitude lower than $T_{BG}$, and the observed slope of $A_{T,1} \sim 0.25$ kΩ/K is much higher than expected from an electron-phonon mechanism. Since $T_{BG} \propto \sqrt{(n)}$ [25], it has been suggested that near the charge neutrality point ($\nu = 0$) or at the Fermi energy resets ($\nu = \pm 2$) [5], the values $n$ and $T_{BG}$ become small enough to yield a $\rho_{e-ph} \propto T$ dependence with an enhanced slope that can persist to temperatures as low as $T = 0.5$ K [13, 26]. In addition to the temperature range, this scenario however appears inconsistent with our findings as we observe 1) a large interval of dopings where the low-$T$ linear regime occurs, 2) the emergence of a $T^2$-dependent resistivity of the same amplitude near the flat band edges and 3) the evolution of the prefactor $A_{T,\gamma}$ as a function of doping, which sharply increases for $\nu < -3$, as shown in the inset of Fig.1.d.

These low-temperature observations are summarized in a *schematic* phase diagram in Fig.2.a. The metallic ground state at the edges of the band displays typical Fermi liquid behavior: a quadratic $T^2$-dependent resistivity. However, doping away from the band edges induces 'zero-temperature' phase transitions, to a non-Fermi liquid state with a $T$-linear resistivity for filling factors $-3.5 < \nu < -2$. It is thus tempting to identify the latter state as a 'strange' metal, wherein the finite-$T$ metallic properties are dominated by critical fluctuations [15-20]. Support for this interpretation is offered by a comparison to the celebrated property of strange metals - a quasiparticle scattering time which is set by the Planckian limit $1/\tau = k_B T/\hbar$ regardless of the nature of the scattering events [30]. A rough estimate of the scattering rate (see SI section D) indicates that the inelastic $T$-linear resistivity is indeed consistent with Planckian dissipation, confirming previous reports [12]. Overall these findings resemble the $T$-linear resistivity observed in LSCO and Bi2201 [31] down to $T \to 0$, both at and away from a quantum critical point, and are possibly associated with the reconstruction of the Fermi surface [32].

Additional evidence for the quantum-critical scenario is provided by the superconducting state close to $\nu \approx -2$, where we observe a recovery of the strange-metal phase upon suppression of the superconducting dome by a small perpendicular magnetic field $B_c = 300$ mT (Fig. 2.b). As shown in Fig.2.c at 40 mK, above $B_c$, the magnetoresistivity (MR) scales linearly $\rho(B) \propto B$ up to $B = 1.5$ T. This allows us to evaluate the MR-corrected resistivity of the superconductivity-suppressed state from the zero-field intercept of the linear $B$-dependence. Following this protocol for all temperatures we find similar 'strange' metal behavior $\rho(T) = \rho_0 + A_{T,1}T$ with a slope of $A_{T,1} \sim 0.3$ kΩ/K, for $T < T_c$ as for $T > T_c$ (Fig.2.b). A similar universal recovery of the 'strange' metal phase below the superconducting dome was previously reported in a wide variety of strongly-correlated systems [20, 31], in spite of the vast differences in their Fermi surfaces, their structural and magnetic properties.

Interestingly, the $T$-linear resistivity is accompanied by a $B$-linear magneto-resistivity (MR) also outside the superconducting dome. Such linear MR provides additional evidence for the existence of critical fluctuations that interact with the metallic ground state, and are characteristic of 'strange' metal phases in a multitude of superconductors [33, 34]. We illustrate the evolution of the MR across the band, where in Fig.3.a we show the (numerical) derivative $\left( \partial \rho / \partial B \right)_\nu$ and in Fig.3.b the resistivity vs. magnetic field for a variety of fillings. To quantitatively analyze the power-law scaling of the MR, we use a similar analysis as for the *T*-dependent resistivity, where we assume a *B*-dependence $\rho(B) = \rho_0 + A_{B,\gamma}B^\gamma$, where the parameters $A_{B,\gamma}$ and $\gamma$ define the pre-factor and the exponent of the power-law *B*-dependence. We fit each *B*-field segment of the curves in Fig.3.b with $A_{B,\gamma}B^\gamma$, and plot the evolution of the corresponding logarithmic derivative $\gamma = \partial[ln(\rho(B) - \rho_0)]/\partial[ln(B)]$ as a function of *B* in Fig. 3.c. Overall we find a behavior very similar to the temperature dependence. In proximity of the charge neutrality point ($\nu = -0.2$) and of full filling ($\nu = -3.7$), we find a $\gamma \approx 2$ and a $\rho(T) \propto B^2$ dependence for low B-fields B < 0.4T, which saturates at higher field. This saturation, as well as the saturation of the resistivity at high temperature are discussed in the SI (section F). For the filling factor range of $-3.5 < \nu < -2$ we find a $\gamma \sim 1$, which gives rise to a linear $\rho(T) \propto B$ dependence with an ultra-high slope of $A_{B,1} > 2$kΩ/K for $B > 0.2$ T (shown in the inset of Fig.1.d), but which saturates for $B \to 0$. We additionally show similar concomitant *B*-linear (up to 3T) and *T*-linear resistivities in another device with nearly identical twist angle (only 0.01° apart) in the section H of the SI. Similar study of the electron-doped region, although made more difficult by the presence of successive phase transitions, also highlights the emergence of a (*B*, *T*)-linear resistivity in the center of the electron-doped flat band (see SI section I).

The observed linear MR is distinct from previous reports of linear MR for graphene systems which originate from classical effects [35] and persist to *T* = 300K and *B* = 62 T [36]. Our MR is ≈100 times stronger and is considerably more fragile, as it saturates near 1T and is quickly suppressed at elevated *T*, as is illustrated in Fig. 3.d. As can be seen in Fig. 3.e, between 40 mK and 50 K the slope of the MR ($A_{B,1}$) decreases ten-fold. In contrast, for devices with $\theta > 1.3°$, we observe an almost absent magneto-resistivity (see SI section C). The identical scaling for the *B*- and *T*-dependences suggests the absence of an intrinsic energy scale in the ground state. We propose that it is dominated by scattering off quantum fluctuations, with a quasiparticle scattering rate which is given by the dominant energy scale $\hbar/\tau = \max(\hbar/\tau_B, \hbar/\tau_T) = \max(\beta\mu_B B, \alpha k_B T)$, where $\tau_B$ and $\tau_T$ are the magnetic and thermal scattering times, and $\mu_B$ is the Bohr magneton $\alpha$ and $\beta$ are numerical factors. We estimate that at 1.9 K, $\tau_B = \tau_T$ at *B* = 0.25 T (see SI section F for a detailed discussion), which corresponds exactly to the onset of the saturation observed in the low-field MR in Fig. 3.c. Hence, for $B \to 0$ we can conclude that the MR saturates because of finite temperature effects, similarly to the case of LSCO [34]. The Ansatz proposed to describe transport relaxation rates of most pnictides, cuprates, heavy fermions and twisted chalcogenides near a quantum critical point, $\hbar/\tau = \sqrt{(\alpha k_B T)^2 + (\beta\mu_B B)^2}$ [20, 33], does not account for the MR of MATBG at finite temperatures.

These findings make a clear case that MATBG hosts a Planckian-limited *T*-linear resistivity that extends down to unprecedently low temperatures of 40 mK and occurs alongside a quantum *B*-linear magnetoresistance. Such behavior is incompatible with a purely Fermi-liquid picture with conventional electron-phonon scattering. Nonetheless, Fermi-liquid behavior is observed throughout the entire moiré band at non-magic angles (as shown in a study of a variety of twist angles discussed in section B of the SI); in contrast, in MATBG it is pushed to the flat-band edges. We therefore conclude on the existence of a 'strange-metal' phase arising from a quantum-critical region spanning a range of dopings including but not limited to those where the Fermi surface reconstructs, and where quantum fluctuations dominate the metallic ground state of MATBG. Our observation reveal that superconductivity in MATBG emerges in a state dominated by quantum fluctuations. While the precise relationship between quantum fluctuations and phase transitions, as well as the microscopic nature of the fluctuations, remains an open problem, our work establishes a clear connection between strongly correlated and highly-tunable electronic moiré systems and the universality class of the quantum critical matter.

**Acknowledgements:** We are grateful for fruitful discussions with Allan MacDonald, Pablo Jarillo-Herrero and Piers Coleman. D.K.E. acknowledges support from the Ministry of Economy and Competitiveness of Spain through the "Severo Ochoa" program for Centres of Excellence in R&D (SE5-0522), Fundació Privada Cellex, Fundació Privada Mir-Puig, the Generalitat de Catalunya through the CERCA program and funding from the European Research Council (ERC) under the European Union's Horizon 2020 research and innovation programme (grant agreement No. 852927)". J.D.M. acknowledges support from the INphINIT 'la Caixa' Foundation (ID 100010434) fellowship programme (LCF/BQ/DI19/11730021). G.d.B. acknowledges funding from the "Presidencia de la Agencia Estatal de Investigación" within the "Convocatoria de tramitación anticipada, correspondiente al año 2019, de las ayudas para contratos predoctorales (Ref. PRE2019-088487) para la formación de doctores contemplada en el Subprograma Estatal de Formación del Programa Estatal de Promoción del Talento y su Empleabilidad en I+D+i, en el marco del Plan Estatal de Investigación Científica y Técnica y de Innovación 2017-2020, cofinanciado por el Fondo Social Europeo. I.D. acknowledges the support from INphINIT "La Caixa" (ID 100010434) fellowship program (LCF/BQ/DI19/11730030). K.W. and T.T. acknowledge support from the Elemental Strategy Initiative conducted by the MEXT, Japan (grant number JPMXP0112101001) and JSPS KAKENHI (grant numbers 19H05790 and JP20H00354). L.L. acknowledges support from the Science and Technology Center for Integrated Quantum Materials, NSF Grant No. DMR-1231319; and Army Research Office Grant W911NF-18-1-0116.

**Author contributions:** D.K.E. and X.L. conceived and designed the experiments; I.D., G.d.B., J.D-M. and X.L. fabricated the devices; I. D., A. J. G.d.B., J.D-M. and X.L. performed the measurements; A.J. analyzed the data; A.J., H.I. and L.L. performed the theoretical modeling; T.T. and K.W. contributed materials; D.K.E. supported the experiments: A.J. and D.K.E. wrote the paper.

**Competing financial and non-Financial interests:** The authors declare no competing financial and non-financial interests.

**Figure :**

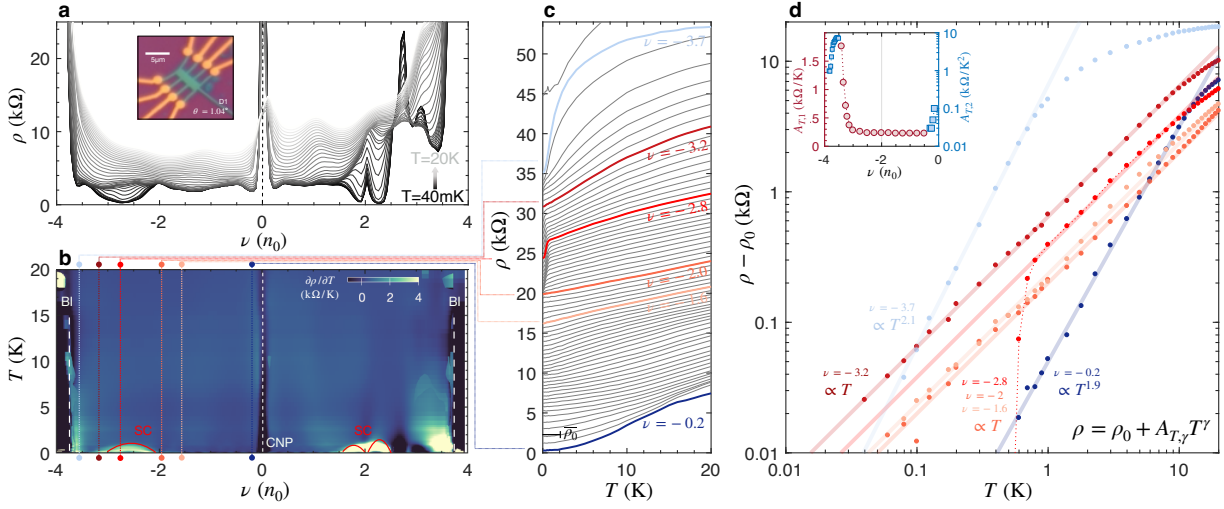

**Fig. 1. Temperature dependence of the resistivity of hole-doped MATBG. a** Resistivity $\rho$ of device D1 ($\theta = 1.04°$) as a function of the filling factor $\nu$ for successive temperatures ranging from 40 mK to 20K. The charge neutrality point (CNP), band insulators (BI) and the superconducting (SC) domes are highlighted. Away from these regions, a metallic ground state is observed. Inset shows an optical image of the Hall bar device in which a four-terminal geometry is used to measure $\rho$. **b** Map of $\partial\rho/\partial T$ vs. $(\nu, T)$ in D1. Vertical dotted lines represent the line cuts which are highlighted in the next sub-figure. **c** Temperature dependence of the resistivity for successive filling of the flat band from near CNP (bottom) to near full-filling (top). Curves are shifted for clarity and the typical residual $T \to 0$ resistivity $\overline{\rho_0}$ is shown. The low-$T$ resistivity evolves from a super-linear $T$-dependence at the band edges, to a $T$-linear dependence in its center. **d** Log-log plot $\rho(T) - \rho_0$ vs. $T$ for the highlighted filling factors in c. Power law fits of the low-$T$ dependence are shown by straight lines. An evolution from a quadratic $\rho \propto T^2$ dependence near CNP and full-filling, to a linear $\rho \propto T$ dependence inside the flat-band region is seen. Inset shows the evolution of the prefactor $A_{T,\gamma}$ upon doping both for the linear-in-$T$ resistivity (left y-axis) and the $T$-square resistivity (right y-axis).

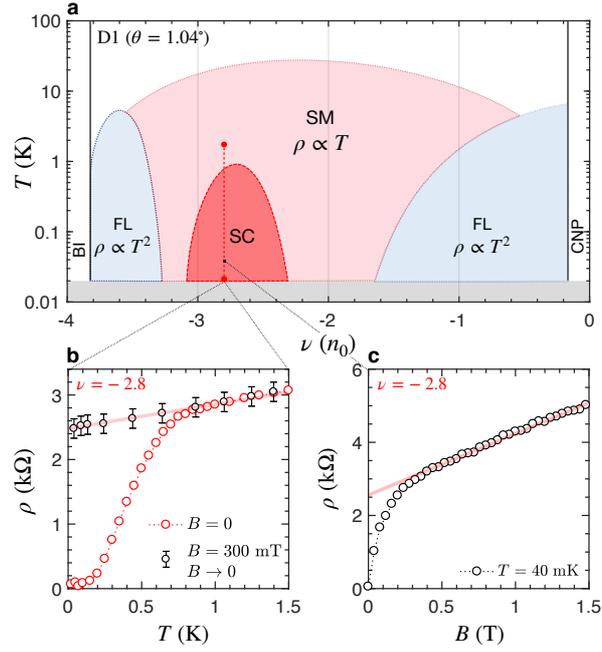

**Fig. 2. Quantum critical fan. a** Schematic representation of the ($\nu,T$) phase diagram of hole-doped MATBG. The superconducting dome is enclosed in a 'strange' metal region dominated by quantum fluctuations. The canonical Fermi liquid behavior is recovered near the boundary of the flat-band region. **b** $\rho(T)$ for $B = 0$ across the superconducting phase transition at $\nu = -2.8$, and the in-field corrected $\rho(T)$ for the critical field $B_c = 300$mT. After suppression of the superconducting order, the uncovered metallic state is a 'strange' metal. Error bars correspond to the 95% confidence bound on the linear fit of the magnetoresistivity. **c** Evolution of the resistivity at $\nu = -2.8$ and $T = 40$mK vs. $B$. The suppression of the superconducting order leads to a sharp increase of the resistivity, and is followed by a linear MR up to $B = 1.5$ T. The linear MR is highlighted by a solid red line.

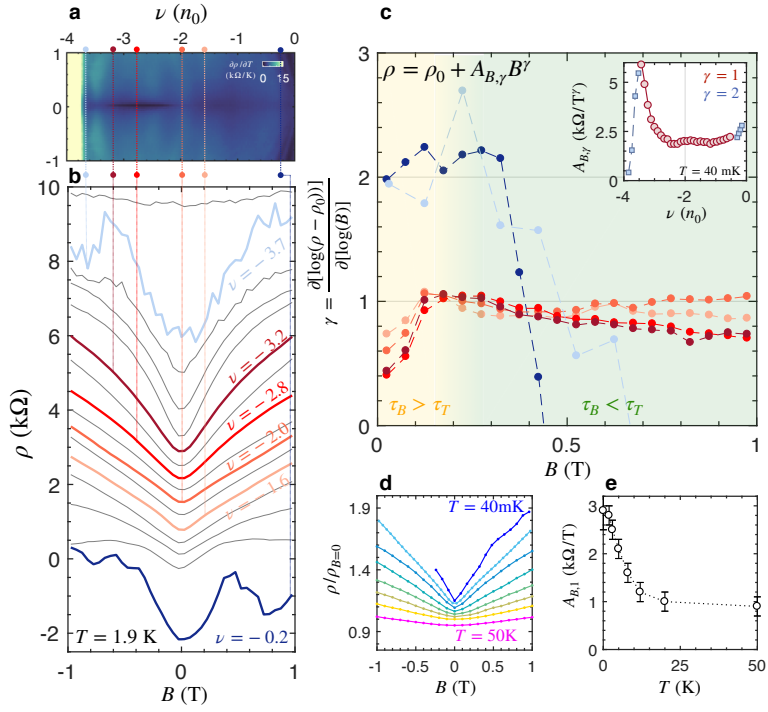

**Fig. 3. Magnetic field dependence of the resistivity of hole-doped MATBG. a** Map of $\partial \rho/\partial B$ as a function of $(\nu, B)$. **b** Resistivity $\rho$ as a function of magnetic field $B$ for several filling factors (matching those detailed for the $T$-dependence in Fig.1). **c** Shows the logarithmic derivative $\gamma = \partial[ln(\rho(B) - \rho_0)]/\partial[ln(B)]$ vs. $B$ for the highlighted filling factors. A $B$-linear MR is found in the 'strange' metal phase in the band center, while a $B^2$-dependence prevails at the band edges. A saturation is seen at low field B < 0.2 T, which marks the crossover from thermal and magnetic dominated scattering times $\tau_T = \tau_B$. Inset shows the evolution of the linear and quadratic prefactors $A_{B,\gamma}$ upon doping. **d** Plots of $\rho(B)/\rho_0$ vs. $B$ for different temperatures from $T$ = 40 mK to $T$ = 50 K. Curves are shifted for clarity. The linear magnetoresistance is suppressed upon increasing temperature. **e** Shows the slopes $A_{B,1}$ as extracted from d, as a function of $T$. Error bars correspond to the 95% confidence bound on the magnetoresistivity linear fit.

**References:**

[1] R. Bistritzer &. A. H. MacDonald. Moiré bands in twisted double-layer graphene. Proceedings of the National Academy of Sciences 108, 12233-12237 (2011).
[2] Y. Cao et al. Correlated insulator behaviour at half-filling in magic-angle graphene superlattices. Nature 556, 80-84 (2018).
[3] H. C. Po et al. Origin of Mott insulating behavior and superconductivity in twisted bilayer graphene. Physical Review X 8, 031089 (2018).
[4] X. Lu et al. Superconductors, orbital magnets and correlated states in magic-angle bilayer graphene. Nature 574, 653-657 (2019).
[5] U. Zondiner et al. Cascade of phase transitions and Dirac revivals in magic-angle graphene. Nature 582, 203-208 (2020).
[6] D. Wong et al. Cascade of electronic transitions in magic-angle twisted bilayer graphene. Nature 582, 198-202 (2020).
[7] Y. Cao, et al. Unconventional superconductivity in magic-angle graphene superlattices. Nature 556, 43-50 (2018).
[8] M. Yankowitz et al. Tuning superconductivity in twisted bilayer graphene. Science 363, 1059-1064 (2019).
[9] P. Stepanov et al. Untying the insulating and superconducting orders in magic-angle graphene. Nature 583, 375-378 (2020).
[10] Y. Saito et al. Independent superconductors and correlated insulators in twisted bilayer graphene. Nature Physics 16, 926-930 (2020).
[11] A. L. Sharpe, E. J. Fox, et al. Emergent ferromagnetism near three-quarters filling in twisted bilayer graphene. Science 365, 605-608 (2019).
[12] Y. Cao et al. Strange metal in magic-angle graphene with near Planckian dissipation. Physical Review Letters 124, 076801 (2020).
[13] H. Polshyn et al. Large linear-in-temperature resistivity in twisted bilayer graphene. Nature Physics 15, 1011-1016 (2019).
[14] A. A. Abrikosov & I. M. Khalatnikov. The theory of a Fermi liquid (the properties of liquid $^3$He at low temperatures. Reports on Progress in Physics 22, 329 (1959).
[15] C. Proust &. L. Taillefer. The remarkable underlying ground states of cuprate superconductors. Annual Review of Condensed Matter Physics 10, 409-429 (2019).
[16] R. L. Greene, P. R. Mandal, N. R. Poniatowski & T. Sarkar, T. The strange metal state of the electron-doped cuprates. Annual Review of Condensed Matter Physics, 11, 213-229 (2020).
[17] S. A. Grigera et al., Magnetic field-tuned quantum criticality in the metallic ruthenate $Sr_3Ru_2O_7$. Science 294, 329-332 (2001).
[18] T. Shibauchi et al., A quantum critical point lying beneath the superconducting dome in iron pnictides. Annual Review of Condensed Matter Physics 5, 113-135 (2014).
[19] H. V. Löhneysen et al. Non-Fermi-liquid behavior in a heavy-fermion alloy at a magnetic instability. Physical Review Letters 72, 3262 (1994).
[20] A. Ghiotto et al. Quantum criticality in twisted transition metal dichalcogenides. Nature 597, 345-349 (2021).
[21] J. A. N. Bruin, H. Sakai, R. S. Perry & A. P. Mackenzie. Similarity of scattering rates in metals showing T-linear resistivity. Science 339 (6121), 804-807 (2013).
[22] O. Trovarelli et al. $YbRh_2Si_2$: Pronounced non-Fermi-liquid effects above a low-lying magnetic phase transition. Physical Review Letters 85, 626 (2000).
[23] S. Licciardello et al. Electrical resistivity across a nematic quantum critical point. Nature 567, 213-217 (2019).
[24] J. Gonzàlez & T. Stauber. Marginal Fermi liquid in twisted bilayer graphene. Physical Review Letters 124, 186801 (2020).
[25] D. K. Efetov &. P. Kim. Controlling electron-phonon interactions in graphene at ultrahigh carrier densities. Physical Review Letters 105, 256805 (2010).
[26] F. Wu, E. Hwang & S. D. Sarma. Phonon-induced giant linear-in-T resistivity in magic angle twisted bilayer graphene: Ordinary strangeness and exotic superconductivity. Physical Review B 99, 165112 (2019).
[27] X. Liu et al. Tuning electron correlation in magic-angle twisted bilayer graphene using Coulomb screening. Science 371, 1261-1265 (2021).

[28] Y. Saito et al. Isospin Pomeranchuk effect in twisted bilayer graphene. Nature 592, 220-224 (2021).
[29] A. Rozen et al. Entropic evidence for a Pomeranchuk effect in magic-angle graphene. Nature 592, 214-219 (2021).
[30] J. Zaanen. Why the temperature is high. Nature 430, 512–513 (2004).
[31] A. Legros et al. Universal T-linear resistivity and Planckian dissipation in overdoped cuprates. Nature Physics 15, 142-147 (2019).
[32] R. Daou et al. Linear temperature dependence of resistivity and change in the Fermi surface at the pseudogap critical point of a high-$T_c$ superconductor. Nature Physics 5, 31-34 (2009).
[33] I. M. Hayes et al. Scaling between magnetic field and temperature in the high-temperature superconductor $BaFe_2(As_{1-x}P_x)_2$. Nature Physics 12, 916-919 (2016).
[34] P. Giraldo-Gallo, J. A. Galvis, et al. Scale-invariant magnetoresistance in a cuprate superconductor. Science 361, 479-481 (2019).
[35] Z. M. Liao et al. Observation of both classical and quantum magnetoresistance in bilayer graphene. Europhysics Letters 94, 57004 (2011).
[36] F. Kisslinger et al. Linear magnetoresistance in mosaic-like bilayer graphene. Nature Physics 11, 650-653 (2015).

**Methods:**

**Supplementary Information** is available for this paper.

**Correspondence and requests for materials** should be addressed to D.K.E.

**Data availability statement:** Source data are provided with this paper. All other datasets that supports the plots within this publication are available from the corresponding authors upon reasonable request.

**Reprints and permissions information** are available.

**Methods section:**

Screening layer fabrication process: The devices presented in this study were produced following the "cut-and-stack" method: a thin hBN flake is picked up with a propylene carbonate (PC) film, then placed on a 90 ºC polydimethyl siloxane (PDMS) stamp. The hBN flake then allows to pick up a portion of a pre-cut monolayer graphene flake previously exfoliated mechanically on $Si^{++}/SiO_2$ (285nm) surface. Then, the remaining graphene sheet is rotated to a target angle usually around 1.1º - 1.15º and picked up by a hBN/graphene stack on PC (introduced previously). The resulting heterostructure is placed on top of another thin hBN flake (which thickness is chosen by optical contrast and further confirmed with atomic force microscopy measurements). Finally, the last layer of the heterostructure, sitting at the very bottom, is composed of a graphite flake (typically a few layer graphene thick) to create both a local back gate and a screening layer. The final stack is then placed on a target $Si^{++}/SiO_2$ (285 nm) wafer, where it is etched into a multiple Hall bar geometry using $CHF_3/O_2$ plasma and edge-coupled to Cr/Au (5/50 nm) metal contacts.

Transport measurements: Transport measurements were carried out in a dilution refrigerator (with base temperature 20-40mK) and an "Ice Oxford" VTI (with base temperature 1.6K). We used a standard low frequency lock-in technique using Stanford Research SR860 amplifiers with excitation frequency $f$ = 13.131 Hz and Stanford Research SR560 pre-amplifiers. The back-gate voltage was controlled from Keithley 2400s voltage source-meters. DC voltage vs. DC excitation current measurements are also performed using SR560 low-noise DC voltage preamplifier in combination with a Keithley 2700 multimeter.

Superconducting-type coaxial cables (2m long, Lakeshore) connect the mixing chamber plate to the room-temperature plate. Each line is equipped with a pi filter (RS-239-191) at room temperature, a powder filter (Leiden cryogenics) and a two-stage resistor-capacitor filter at the mixing chamber plate to avoid any unwanted heating of the charge carriers at low temperature. Our measurements show no sign of saturation of the electronic temperature down to 40mK. Investigations to lower temperatures will require advanced filtering to avoid heating of the charge carriers and advanced thermometry techniques to accurately evaluate the electronic temperature. The measurement were realized for various excitation currents $I$ (obtained by applying a tension to a 10 MΩ resistor) ranging from ($I$<10 nA) not to break the superconducting state and up to $I$=200 nA (to enhance the signal in the metallic phase), with a special attention given to ensure consistency between datasets and not to increase the electronic temperature by gradually increasing the current and tracking the output voltage. All in-field measurements reported in this work were realized with an applied out-of-plane magnetic field.

# Supplementary Information: 'Quantum critical behavior in magic-angle twisted bilayer graphene'

Alexandre Jaoui[1,*], Ipsita Das[1], Giorgio Di Battista[1], Jaime Díez-Mérida[1], Xiaobo Lu[1], Kenji Watanabe[2], Takashi Taniguchi[2], Hiroaki Ishizuka[3,4], Leonid Levitov[4] and Dmitri K. Efetov[1,*]

1. ICFO - Institut de Ciencies Fotoniques, The Barcelona Institute of Science and Technology, Castelldefels, Barcelona, 08860, Spain.
2. National Institute for Material Sciences, 1-1 Namiki, Tsukuba, 305-0044, Japan.
3. Department of Physics, Tokyo Institute of Technology, Meguro, Tokyo, 152-8551, Japan.
4. Department of Physics, Massachusetts Institute of Technology, Cambridge MA 02139, USA.

*Correspondence to: alexandre.jaoui@icfo.eu and dmitri.efetov@icfo.eu.

## A. Table of studied devices

| Device | Twist angle $\theta$ (°) | Width x Length (μm$^2$) |
|---|---|---|
| D1 | 1.04 | 3x2 |
| D2 | 1.10 | 2.5x3 |
| D3 | 1.03 | 2.9x3.6 |
| D4 | 1.02 | - |
| D5 | 1.3 | 3x3.5 |
| D6 | 1.4 | 1.8x2.8 |
| D7 | 1.5 | 1.4x7.4 |
| D8 | 1.05 | 1.8x10.4 |

**Supplementary Table S1:** Geometry of the various devices discussed in this study.

## B. Extended temperature dependent data of studied devices

Supplementary Fig.S1 shows the resistivity vs. temperature for several devices with various twist angles around the magic angle, D2 ($\theta = 1.10°$), D3 ($\theta = 1.03°$), D4 ($\theta = 1.02°$, D5 ($\theta = 1.3°$) and D7 ($\theta = 1.5°$) respectively in subplots a, b, c, d and e. The dimensions of the devices are presented in Supplementary Table S1. Devices D2 and D3 both show a succession of low-temperature superconducting and correlated-insulating states in the center of the flat-band region. These states make the metallic ground state hardly accessible. Yet, one remains able to see in D3 that the resistivity scales super-linearly with temperature near the charge neutrality point and a *T*-linear term develops in the core of the flat band (although here interrupted at low-*T* by several phase transitions). This behavior, described in D1 in the main text, is confirmed in devices D4 and D5 (although not tracked below 1.6 K). In both devices, we observe a clear evolution from a low-temperature superlinear (parabolic) resistivity near charge neutrality into a *T*-linear resistivity extending down to the lowest temperatures in the center of the flat band. With further doping, the slope of this linear resistivity increases until a superlinear resistivity at low temperature is recovered. On the contrary, device D7 with a twist-angle far from the magic angle shows no evidence of low-$T$ linear resistivity. Device D6 shows similar $\rho$ vs. $T$ as device D7: a superlinear (quadratic) temperature dependence across the entire hole-doped flat-band, and is discussed in more detail in Fig. S2. A quantitative study of the power law dependence of resistivity as a function of temperature at half-filling of the hole-doped flat band is featured in Supplementary Fig.S2.

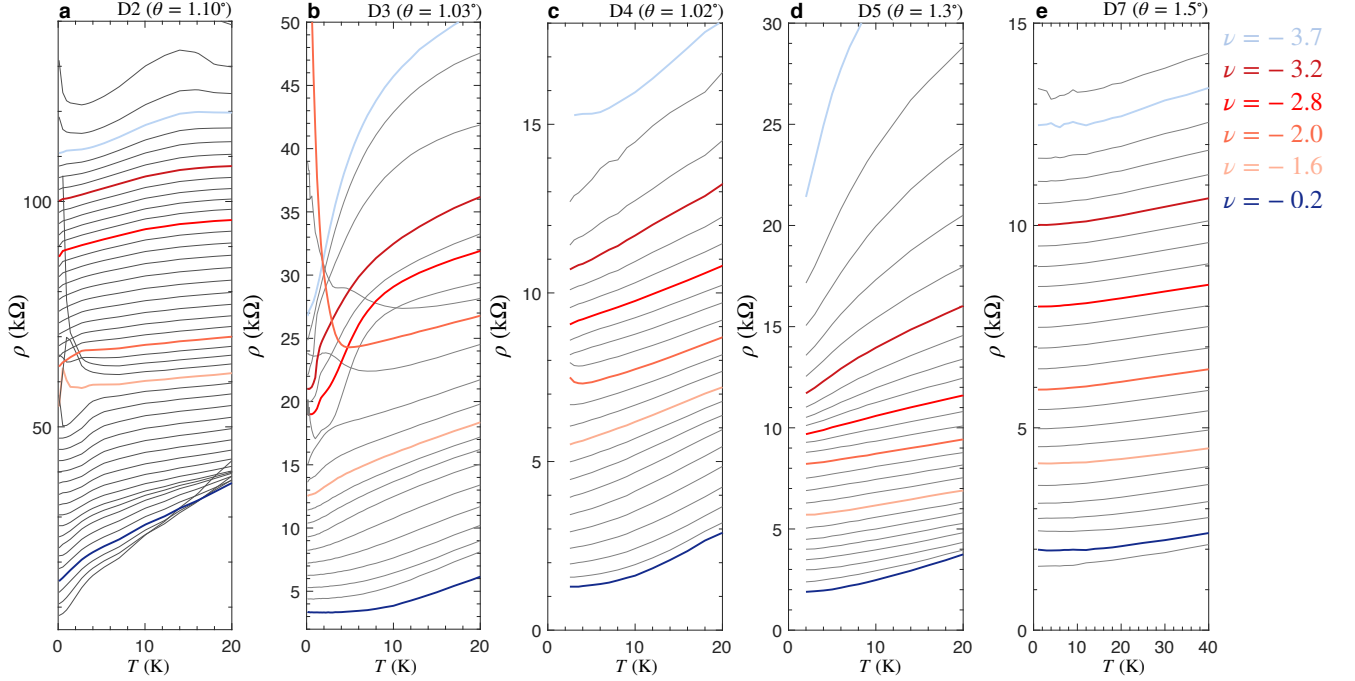

**Supplementary Figure S1:** Evolution of the resistivity of various twisted bilayer graphene devices both close to the magic angle and away from the magic angle. The highlighted filling factors are those discussed on device D1 in the main text. While 'strange' metal behavior, i.e. a linear-in-$T$ resistivity is seen in D3, D4 and D5, device D7 shows only superlinear (parabolic) temperature dependence across the entire flat band.

## C. Vanishing 'strange' metal phase in non-magic-angle devices.

Our report makes a convincing case that strong electronic correlations in the flat bands of MATBG drive the emergence of the low-temperature linear resistivity in quantum-critical scenario. To check this intricate connection, we explore in more detail other devices with twist angles that deviate from the magic angle (presented above in Supplementary table S1). These devices have much weaker or absent correlations due to strongly increased bandwidths. We find that the transport properties of such devices do not show signs of quantum-critical behavior, as it is highlighted in Fig. S2.a and b. Both figures examine the temperature and field dependences of device D6 with a twist angle of $\theta = 1.4°$. These show $\rho(T)$ and $\rho(B)$ traces for the same filling factors as previously discussed for device D1. In stark contrast to device D1, we do not observe any sign of a linear-in-$T$ and linear-in-$B$ resistivity extending down to the lowest excitation (as stated in the manuscript, and well described in the literature [2,14], above the Bloch-Gruneisen temperature a $T$-linear resistivity is expected). Instead, we find a parabolic temperature dependence across the entire hole-doped band, and an almost absent magnetoresistivity, which only shows a very weak and superlinear dependence for $\nu < -2.8$. To highlight the dependence of quantum critical transport on the strength of the correlations, we examine a variety of devices, D1 to D7, with different twist angles from $\theta = 1.02°$ to $1.5°$ in Fig. S2.c. Similarly to what we have reported in Fig.1.d. of the main text, the resistivity of all devices is shown on a log-log scale as $(\rho(T) - \rho_0)$ for a filling of $\nu = -2$. We notice that devices D1 to D5, which are close to the magic angle, all show linear-in-$T$ resistivity below $T < 20$ K.

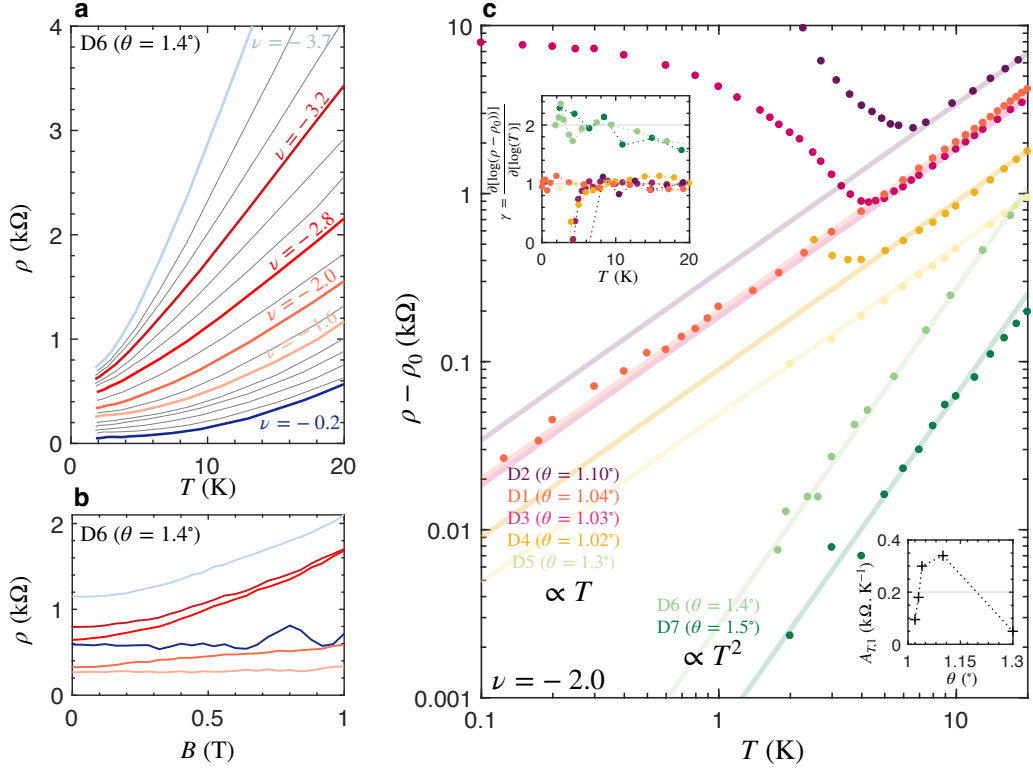

**Supplementary Figure S2: a** Temperature dependence of the resistivity of device D6 ($\theta = 1.4°$) for the filling factors discussed for device D1 in Fig.1 and Fig.3 of the manuscript. The low-temperature dependence remains superlinear over the whole flat-band region. **b** Magnetoresistivity of device D6 for the aforementioned filling factors. The MR vanishes or shows weak and non-linear B-dependence, in stark contrast with that of device D1. **c** Log-log graph of $(\rho(T) - \rho_0)$ as a function of $T$ for devices D1 to D7 (sorted by increasing twist angle) at filling $\nu = -2$. The devices are presented in the supplementary table 1. Devices D1 to D5 display $T$-linear resistivity below $T < 20$ K. The specificity of device D1, i.e. its correlated insulating state is extinct due to screening, is illustrated here. On the contrary, devices D6 and D7 with twist angles the furthest away from the magic angle, show a low-$T$ $T^2$-dependent resistivity. Top inset underlines the power law dependence from the temperature logarithmic derivative $\gamma = \partial\left[ln(\rho(T) - \rho_0)\right]/\partial\left[ln(T)\right]$. The evolution of the slope $A_{T,1}$ as a function of twist angle is shown in the bottom inset.

While this resistivity scales linearly down to the lowest temperatures $T = 40$ mK in D1, it is interrupted by a metal-insulator transition in most other devices. The evolution of the half-filling slope $A_{T,1}$ with the twist angle is shown in the bottom inset of figure S2.c. $A_{T,1}$ peaks around the magic angle $\theta = 1.10°$ at a value of $0.35$ k$\Omega$/K. This angle-dependent slope is in agreement with previous reports [1,2]. On the other hand, the devices twisted further away from the magic angle show a $T^2$-dependence, i.e. typical Fermi liquid behavior, across their entire band at low temperatures. This evolution of the low-temperature dependence of the resistivity at half-filling $\nu = -2$, from $T$-linear in devices close to the magic angle to a $T^2$-dependence at larger twist angles is further clarified by the evolution of the logarithmic derivative of the inelastic resistivity $\gamma = \partial\left[ln(\rho - \rho_0)\right]/\partial\left[ln(T)\right]$ as a function of temperature in the top inset of Fig. S2.c. Additionally, the prefactor of the $T^2$ term (putative y-axis intercept in a log-log scale) decreases as the device is twisted away from the magic angle. In the Fermi liquid picture, $\rho(T) = \rho_0 + A_{T,2}T^2$ and $A_{T,2} \propto 1/E_F^2$ (as well as the electronic specific heat) are enhanced by electronic correlations, as described though the Kadowaki-Woods ratio [3, 4]. This suggests that electronic correlations in the Fermi-liquid ground state are also enhanced near the magic angle. Although elusive in most graphene systems, a $T^2$-dependent resistivity associated with Umklapp scattering has recently been reported in graphene-hBN superlattices [5].

### D. Planckian dissipation

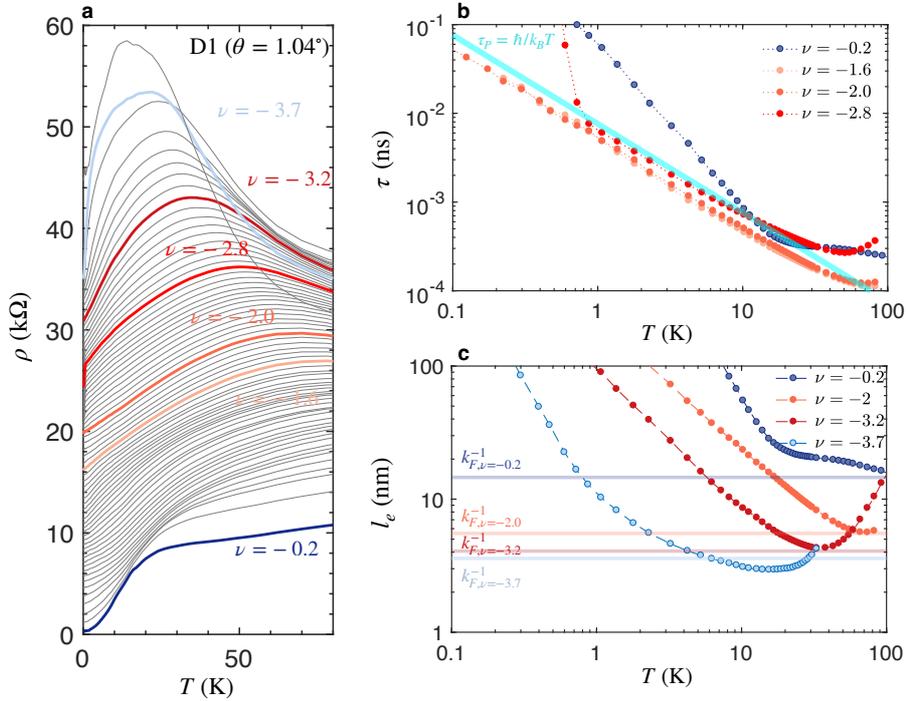

**Supplementary Figure S3: a** Evolution of the resistivity of device D1 up to 80 K. **b** Evolution of the inelastic scattering time with temperature as evaluated from the Drude formula for the highlighted filling factors at which we have determined the effective mass of the carriers *m\** and the carrier density *n* from a study of the Shubnikov de Haas effect (in device D8, see below). The inelastic scattering rate associated with the *T*-linear resistivity matches the Planckian scattering rate. A mismatch is expected from our crude approximation of an isotropic Fermi surface which contrasts with the peculiar filling of the flat band [7]. **c** Evolution of the electronic mean free path with temperature. The resistivity appears to saturate when $l_e \sim 1/k_F$ i.e. when the system reaches the Mott-Ioffe-Regel limit.

Supplementary Fig.S3.a shows the resistivity of the device D1, which is extensively discussed in this manuscript, up to $T = 80$ K. As mentioned in the main text, the resistivity displays a maximum at a finite temperature which is shifted to lower *T* with increasing filling of the electronic flat band. This result is in good agreement with previous reports on MATBG, where it was associated with thermal activation of higher energy bands [2].

In the Drude picture, the measured resistivity $\rho$ is related to the scattering time of electrons

($\tau_e$) and holes ($\tau_h$) (with densities $n$ and $p$, respectively) and their masses ($m_e$ and $m_h$, respectively) by Eq.(S1):

$$\rho^{-1} = e^2 \left( \frac{n\tau_e}{m_e} + \frac{p\tau_h}{m_h} \right) \quad (S1)$$

Here, for electrostatic hole doping, we make the assumption that the electronic density is given by the doping $n$. The inelastic scattering time is then given by the equation Eq.(S1) (see Ref [6] for an in-depth discussion on the dichotomy between elastic and inelastic scattering rates in systems bound by Planckian dissipation).

$$\tau = \frac{\left( \frac{1}{\rho - \rho_0} \right) m}{n e^2} \quad (S2)$$

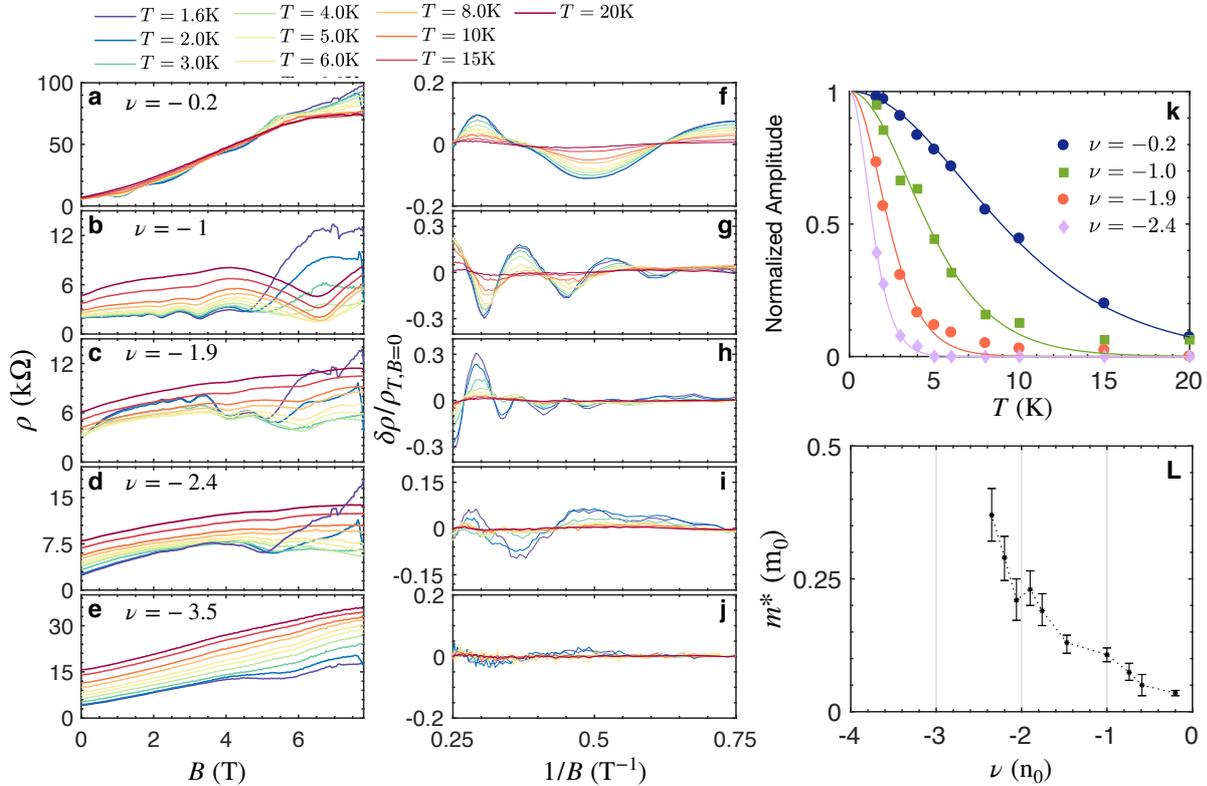

**Supplementary Figure S4: a-e** Evolution of the resistivity $\rho_{xx}$ of device D8 with $B$ for various temperatures and fillings factors. **f-j** Oscillatory part of the magnetoresistance $\delta\rho_{xx}$, obtained by removing a polynomial background to $\rho_{xx}$ shown in subplots a-e. $\delta\rho_{xx}/\rho_{T,B=0}$ is plotted as a function of $1/B$. **k** The amplitude of the $1/B=0.3$ T$^{-1}$ peak is shown as a function of temperature for various filling factors. Fits to the Lifshitz-Kosevich theory, $R_T = X/sh(X)$, where $X = \frac{14.69 T}{B} \frac{m}{m_0}$ where $m_0$ is the bare electron mass are shown as solid line and used to evaluate the quasiparticle effective mass. **L** The effective mass of charge carriers $m$ in device D8 ($\theta = 1.05°$) is shown as a function of the filling factor. The error bars are associated with 95% confidence bounds on the fit.

To evaluate $\tau$ we first need to evaluate the effective mass of the quasiparticles $m$. To do so, we rely on the study of the quantum oscillations of the magnetoresistivity associated with the Shubnikov-de Haas effect. This study could not be carried out in device D1 ($\theta = 1.04°$) which shows no clear quantum oscillations, as seen in Fig.3.a. However, device D8 ($\theta = 1.05°$), with almost identical twist angle, shows a clear set of quantum oscillations across its electronic flat band. This is shown in the evolution of the transverse magnetoresistance $\rho_{xx}$ as a function of the magnetic field at various temperatures for an array of filling factors in D8 in Supplementary Figure S4 subplots (a) to (e). We extract the quantum oscillations by removing a polynomial fit to the total magnetoresistance. The oscillatory part, $\delta\rho_{xx}$, is then plotted as $\dfrac{\delta\rho_{xx}}{\rho_{T,B=0}}$ as a function of $1/B$ for the same temperatures and filling factors in the subplots (f) to (j). Finally, we show the temperature dependence of the amplitude of the $1/B=0.3$ T$^{-1}$ peak for the successive fillings factors in Supplementary Figure S4.k. The effective mass is then evaluated by fitting the temperature-dependent part of the Lifshitz-Kosevich model $R_T = X/sh(X)$, where $X = \dfrac{14.69T}{B}\dfrac{m}{m_0}$ where $m_0$ is the bare electron mass, to this temperature dependence. The fits are shown as solid lines in Supplementary Figure S4.k. The extracted effective mass across the flat band is finally shown in Supplementary Figure S4.l. The error bars are associated with 95% confidence bounds. We observe an increase of $m$ with filling of the flat band from $m = 0.03m_0$ near charge neutrality ($\nu = -0.2$ up to $m = 0.3m_0$ near half-filling ($\nu = -2.5$. The quantum oscillations, however, could not be tracked for $|\nu| > 2.5$. The effective mass we determine across the flat band is in very good agreement with the seminal study of the quantum oscillations in MATBG realized on a device with, again, the same twist angle ($\theta = 1.05°$) [1]. While fine evolutions of the effective mass are expected between devices, the consistency of the effective mass between D8 and the literature justifies the use of effective mass determined in D8 to evaluate the scattering rate in D1 from the (approximate) Drude formula.

We now consider the linear $\rho(T)$ traces from Fig.1.d. of the main manuscript where we find ultra-steep slopes with $A_{T,1} = 0.3$kΩ/K for most filling factors. Fig. S3.b. shows the extracted Drude inelastic scattering times vs. temperature for several filling factors, and compares these with the temperature dependence of the Planckian scattering time. We find that $\tau \sim \hbar/k_BT$ for most filling factors where $\rho(T) \propto T$, which firmly confirms that the 'strange' metal phase of MATBG is set by the universal Planckian limit [1]. As aforementioned, discrepancies can be explained both from the Drude formula (in light of the peculiar fermiology of MATBG [7]) and from the evaluation of the effective mass. Also consistent with this finding, we find that for filling factors where a Fermi liquid behavior prevails, i.e. $\nu = -0.2$, the scattering times are larger than the Planckian limit, $\tau > \hbar/k_BT$. Interestingly, for one filling factor, $\nu = -3.5$, we find a skyrocketing slope $A_{T,1}$ which is enhanced almost tenfold (as can be seen in the inset of Fig.1.d). While for this filling $m$ could not be directly determined due to the absence of clear quantum oscillations, upon assuming that the scattering rate remains Planckian, the increase of $A_{T,1}$ translates into a steep increase of $m \propto A_{T,1}$. Such a sharp enhancement of the effective mass in the direct vicinity of a QCP has been previously reported in $f$-electron systems and calls for further investigation [8].

**E. Saturation of resistivity at the Mott-Ioffe-Regel limit**

Similarly to the scattering rate, the Drude model allows us to evaluate the inelastic mean-free-path $l_e$. In the degenerate regime, $T < T_F$, given by the Eq. (S3) and shown in Fig.S3.c vs. temperature.

$$l_e = \left(\frac{\hbar}{e^2}\right)\left(\frac{1}{\rho - \rho_0}\right)\sqrt{\frac{\pi}{n}} \qquad (S3)$$

The Mott-Ioffe-Regel limit corresponds to the crossover from coherent to incoherent quasiparticle transport and is reached when the charge carrier's mean-free-path falls below any conceivable length scale of the system ($l_e < a, k_F^{-1}$) where $a$ is the lattice constant. Our investigation of MATBG are limited to doping levels such that $k_F^{-1} > a$. We find that $l_e$ decreases with temperature and saturates at $1/k_F$. We argue that the resistivity saturates when the electronic mean-free-path becomes comparable with the Fermi wavelength $l_e \sim 1/k_F$, i.e. upon reaching the Mott-Ioffe-Regel limit which separates coherent and incoherent quasiparticle transport regimes. At even higher temperatures, weakly resistive higher-energy electronic bands become thermally populated [2] and the resistivity decreases. Interestingly, our findings comply with the Mott-Ioffe-Regel bound even though the scattering rate is of the order of the Planckian dissipation rate and the Fermi velocity is small. In contrast, apparent crossing of the Mott-Ioffe-Regel limit associated with non-saturating resistivity have been reported in cuprates [9] and non-degenerate 'strange' metals [10]. This picture may also explain a high field saturation of the magnetoresistivity in Fig. 3.c. of the main text, which is typically found around $B = 1T$. For the case of $B = 1$ T, the value of the magnetic field matches the temperature of 10 K, which is the temperature for which the Mott-Ioffe-Regel limit is reached for $B = 0$. This establishes that the saturation of the resistivity in $(v, B, T)$ is in agreement with $l_e \sim 1/k_F$. Interestingly, a saturation of the linear MR of other 'strange' metals is not observed up to 80T [11,12].

**F. Finite temperature saturation of the magnetoresistivity**

While the similarities in the $T$ and $B$ dependences are striking, there are also subtle differences. The $B$-linear MR does not extend down to the lowest excitations and saturates at finite temperatures below $B < 0.2$ T, which is highlighted in Fig.3.c of the main text. To understand the origin of this saturation, we consider the quasiparticle scattering rate. We define two scattering times: $\tau_B$ defined by $\hbar/\tau_B = \beta\mu_B B$ and its thermal counterpart $\tau_T$: $\hbar/\tau_T = \alpha k_B T$ where $\mu_B$ is the Bohr magneton $\alpha$ and $\beta$ are numerical factors. In the Drude model, the ratio $B/T$ is given by $\tau_T/\tau_B = \beta\mu_B B/\alpha k_B T = A_{B,1}/A_{T,1}(B/T)$. For the case of $v = -2$, and $T = 1.9$ K presented in Fig.3.c, we obtain a slope of $A_{T,1} = 0.26$ kΩ/K and $A_{B,1} = 2.1$ kΩ/T (from Fig.1.d and Fig.3.e respectively). From this we can estimate that at 1.9 K, $\tau_B = \tau_T$ at $B = 0.25$ T, which matches to the onset of the saturation observed in the low field MR, and hence, for $B \to 0$ we can conclude that the MR saturates at low fields because of finite temperature effects.

## G. Evolution from 'strange' metal to Fermi liquid near full-filling

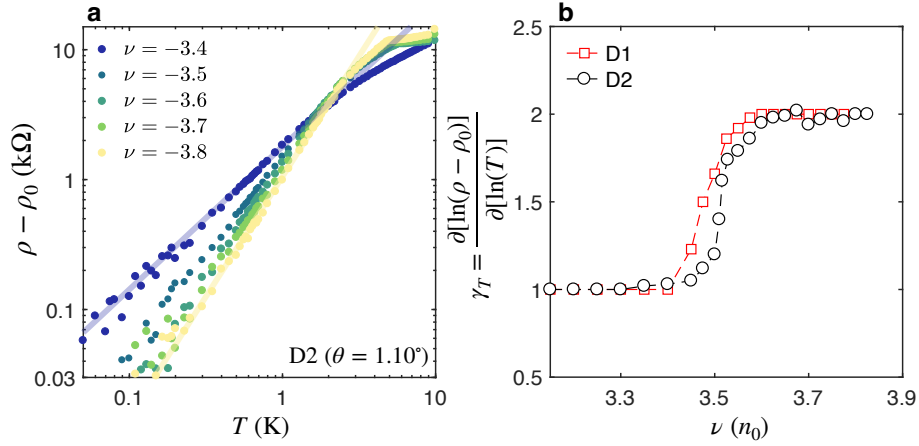

**Supplementary Figure S5: a** Evolution of the resistivity of device D2 ($\theta = 1.10°$) plotted as $\rho(T) - \rho_0$ vs. temperature on a log-log scale for different filling factors. A clear evolution from a $T$-linear to a $T$-quadratic dependence is seen. **b** The logarithmic derivative $\gamma$ as a function of filling factor, shows a clear crossover from $\gamma = 1$ to $\gamma = 2$ near $\nu = -3.5$, for both devices D1 and D2.

Supplementary Fig.S5.a shows the resistivity $\rho(T) - \rho_0$ vs. $T$ on a log-log scale for device D2 with twist angle $\theta = 1.10°$. The residual term $\rho_0$ was determined by fitting the resistivity $\rho = \rho_0 + A_{T,\gamma}T^\gamma$ for $T < 5$ K. The evolution of the slope (in the logarithmic scale) highlights the evolution from $T$-linear to $T$-quadratic upon increasing the filling of the flat band from $\nu = -3.4$ to $\nu = -3.8$. The evolution from the 'strange' metal behavior to a Fermi liquid behavior upon electrostatic doping is further underlined by the study of the evolution of the logarithmic derivative $\gamma = \partial\big[ln(\rho - \rho_0)\big]/\partial\big[ln(T)\big]$ averaged for $T < 5$ K in Fig.S5.b. A clear crossover from $\gamma = 1$ to $\gamma = 2$ near $\nu = -3.5$ is seen in both devices D1 (raw data featured in the main text) and D2. The 'zero'-temperature phase transition from a 'strange' metal to a canonical Fermi liquid is thus seen in both these superconducting devices.

## H. Additional linear magnetoresistance data

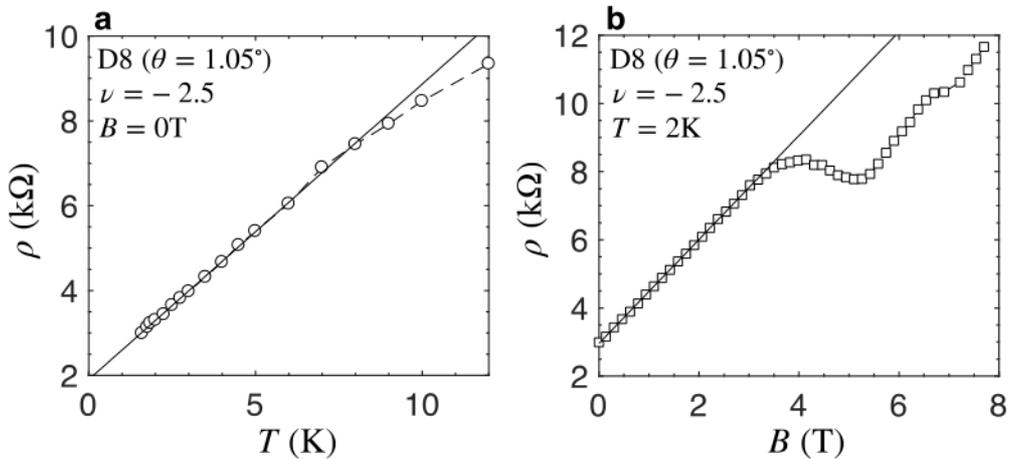

**Supplementary Figure S6: a** Evolution of the resistivity of device D8 ($\theta = 1.05°$) with temperature at $\nu = -2.5$. **b** Evolution of the resistivity of device D8 ($\theta = 1.05°$) with magnetic field at the same filling factor. The $B$-linear resistivity extends up to 3T.

We present both the temperature and magnetic field dependences of the resistivity of device D8 ($\theta = 1.05°$) for $\nu = -2.5$ in Fig.S6.a and S6.b respectively. This device, with an almost identical twist angle as device D1, also shows concomitant linear-in-$T$ and linear-in-$B$ resistivities near half-filling of the flat-band. The $B$-linear resistivity extends here up to 3T.

**I. Quantum critical behavior in electron-doped MATBG device D1**

In the following section, we extend our analysis to the electron-doped device D1. The resistivity as a function of temperature from charge neutrality up to near full-filling ($\nu = +4$) is given in Supplementary Figure S7.a. The temperature dependence shows a similar behavior as that of hole-doped D1: the low-temperature resistivity near the band edges is superlinear. This parabolic resistivity evolves into a linear resistivity at higher temperatures. Starting from the charge neutrality point, the entrance in $T$-linear resistivity is shifted to lower temperature with increasing doping. This linearity could however, not be tracked down to the lowest temperatures, because the region near half-filling hosts two superconducting domes and an insulating state which hide the metallic ground states below roughly 1 K. The superlinear ($T^2$) dependence of the resistivity at low temperature is again recovered near full-filling of the flat band. The evolution from superlinear to linear and vice-versa is highlighted in the log-log plot of ($\rho - \rho_0$) where $\rho = \rho_0 + A_{T,\gamma}T^\gamma$ in the absence of phase transition in Supplementary Figure S7.b. A clear evolution from a Fermi liquid behavior ($T^2$) to a 'strange' metal near half-filling of the electron-doped flat band of device D1 is shown. The resistivity across the superconducting phase transition found at $\nu = +1.8$ (where no insulating state is found) is shown as a function of temperature in Supplementary Figure S7.c and as a function of magnetic field in Supplementary Figure S7.d. Using the linear magnetoresistance, we can extract the resistance of the metallic state obtained by suppressing the superconducting order. This value is featured in Supplementary Figure S7.c and also roughly scales with the expected value from the linear-in-$T$ resistivity above the superconducting transition. This result suggests that upon suppression of the superconducting order, and in the absence of insulating state in the direct vicinity, the 'strange-metal' regime is recovered down to the lowest temperatures. Finally, we turn to the magnetoresistance of electron-doped device D1 shown in Supplementary Figure S7.e. Again, we observed a superlinear magnetoresistance ($B^2$) near the charge neutrality point and full-filling of the flat band, while a linear magnetoresistance emerges near half-filling. This result is, again, similar to that of hole-doped MATBG. Our observations points to the existence of a similar quantum critical region in the center of the electron-doped phase diagram of MATBG, which could, however, not be tracked down to the lowest temperatures because of an array of low-temperature phases transitions.

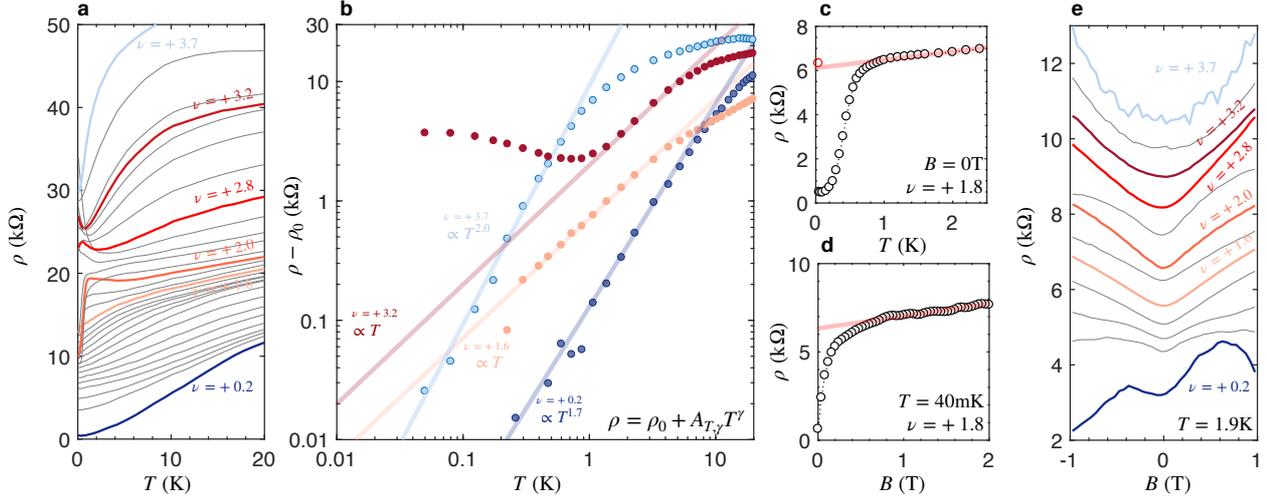

**Supplementary Figure S7: a** Evolution of the resistivity of device D1 ($\theta = 1.04°$) for electron doping from near charge neutrality (bottom) to near full-filling (top) as a function of temperature. Curves are shifted for clarity. **b** Log-log plot of $\rho(T) - \rho_0$ vs. $T$ with $\rho = \rho_0 + A_{T,\gamma} T^\gamma$ for highlighted filling factors. The power law fits of the low-$T$ dependence, in the absence of phase transition, are shown by straight lines. An evolution from a superlinear $\rho \propto T^{\sim 2}$ dependence near charge neutrality and full-filling, to a linear $\rho \propto T$ dependence inside the flat-band region is observed. **c** $\rho(T)$ for $B = 0$ the superconducting phase transition at $\nu = +1.8$, and $\rho(T)$ upon suppression of the superconducting order by a magnetic field $B > B_c$ where $B_c \approx 500$ mT is the critical magnetic field at 40mK (red circle). The uncovered metallic state roughly scales with the linear resistivity of the 'strange' metal phase (red line). **d** Evolution of the resistivity at $\nu = +1.8$ and 40mK vs. $B$ discussed in c. The suppression of the superconducting order leads to a sharp increase of the resistivity, and is followed by a roughly linear MR up to 1 T. The linear MR is highlighted by a solid red line. **e** Resistivity as a function of $B$ for highlighted filling factors. A parabolic magnetoresistivity is observed at the edges of the flat band while a linear magnetoresistivity emerges near half-filling.

## J. A brief survey of theoretical models of electron transport discussed in the main text.

In this section we summarize the results for the theoretical models of transport discussed in the main text, focusing on their predictions for the $T$-dependent resistivity. The predictions of the conventional quasiparticle scattering mechanisms—electron-phonon and electron-electron— strongly differ from the measured $T$ dependence. The Planckian dissipation mechanism, which is a general name for the mechanisms invoking carrier scattering by critical fluctuations, appears to be in a qualitative agreement with the observed linear-$T$ scaling and the measured $d\rho/dT$ values.

As a quick summary, the coupling between **electrons and acoustic phonons** gives rise to a temperature-dependent scattering with a linear $T$ dependence above the Bloch-Gruneisen temperature $T_{BG} = sk_F$, where $s$ is sound velocity and $k_F$ is Fermi momentum. For typical $k_F$ values in MATBG flat bands the temperature $T_{BG}$ is on the order of a few kelvin. At temperatures $T < T_{BG}$ the predicted $T$ dependence of the scattering rate is superlinear, which is unlike the observed $T$ dependence that remains linear down to the temperatures as low as 20 mK. This, along with an abnormally strong slope $d\rho/dT$, excludes the electron-phonon scattering as a mechanism for the observed $T$ dependence.

The scattering mechanism due to **electron-electron interaction in a Fermi liquid** is expected to cause a $T^2$ temperature dependence of scattering, extending throughout a wide temperature range from the lowest temperatures and up to temperatures of the order of the bandwidth. The scattering rate is high due to the uniquely strong e-e interactions in graphene-based systems, and can be further enhanced by the umklapp scattering processes. The $T^2$ scaling matches the $T$-dependent transport observed at doping values near the edges of the flat band, corresponding to a small concentration of electrons or holes. However, it is distinct from the strong linear $T$ dependence $\rho(T)$ observed throughout most of the flat-band range in doping.

These general conclusions are illustrated below for a simple model of a flat band chosen to match the MATBG crystal symmetry and bandwidth, for realistic values of the e-e and e-ph interaction strength.

The scattering mechanism which, according to the reported transport measurements, appears to be at work in MATBG at low temperatures is unique for the **quantum-critical systems.** The properties of the ground state in these systems are dominated by soft modes which are strongly coupled to carriers and produce a $T$-linear scaling of resistivity. This is indeed the scaling observed in MATBG at low temperatures and in a wide range of carrier concentrations. The $T$-linear resistivity was first linked to strong-coupling superconductivity When the high-temperature cuprate superconductors were discovered. Subsequently, $T$-linear resistivity has been seen in the pnictide and organic superconductors, as well as in many heavy fermion compounds, both superconducting and non-superconducting. In most of the heavy fermion materials, the $T$-linear resistivity is seen when they have been tuned by some external parameter to create a low-temperature continuous phase transition known as a quantum critical point. The $T$-linear resistivity is therefore often associated with quantum criticality [23].

Theoretical models explaining the $T$-linear resistivity typically rely on carrier scattering by strongly fluctuating degrees of freedom such as gauge fields [9] or the fluctuations due to the critical soft modes [13]. While presently there is no single universally accepted explanation for the $T$-linear resistivity, it is understood to be a universal property of many strongly-interacting electron systems. The high scattering rate associated with the $T$-linear resistivity is sometimes called "Planckian dissipation", referring to the fact that the scattering rate per kelvin is well approximated by the ratio of the Boltzmann constant to the Planck constant divided by $2\pi$ (e.g. see Ref.[23]).

These values are consistent with those found in the measurements described in the main text.

## A. Resistivity due to electron-phonon scattering

Electron-phonon scattering is often the dominant contribution to the resistivity in metals, in which case the resistivity shows a $\rho \propto T^1$ behavior at $k_B T \gtrsim T_{BG}/4$ whereas $\rho \propto T^4$ at the lower temperatures. This is also likely to be the case of for MATBG [14], but with a smaller energy scale than in typical metals. A recent study also points out that the resistivity shows $\rho \propto T^2$ at a very high temperature where the temperature is comparable or larger than the bandwidth [15], if the nearly-flat band is well separated from the other high-energy bands. To gain ideas on the window of temperature for $T$-linear resistivity by electron-phonon interaction, we have calculated the phonon resistivity using the honeycomb lattice model coupled to acoustic phonons. To see the typical temperatures where the two crossovers takes place, we consider a honeycomb lattice model coupled to acoustic phonons. The electron Hamiltonian reads

$$H_\text{e} = \sum_{\mathbf{k}} \begin{pmatrix} \psi^\dagger_{\vec{k},1} \\ \psi^\dagger_{\vec{k},2} \end{pmatrix}^\text{T} \begin{pmatrix} 0 & h_{\vec{k}} \\ h^*_{\vec{k}} & 0 \end{pmatrix} \begin{pmatrix} \psi_{\vec{k},1} \\ \psi_{\vec{k},2} \end{pmatrix}. \qquad (1)$$

Here $h_{\vec{k}} = t\left(1 + e^{-i\vec{k}\cdot\vec{a}_1} + e^{-i\vec{k}\cdot\vec{a}_2}\right)$ with $t$ the nearest-neighbor hopping, $\vec{a}_{1,2} = a\left(\pm\frac{1}{2}, \frac{\sqrt{3}}{2}\right)$, $|\vec{a}_{1,2}| = a$, are primitive vectors and $\vec{k} = (k_x, k_y)$ is electron momentum. The phonon Hamiltonian

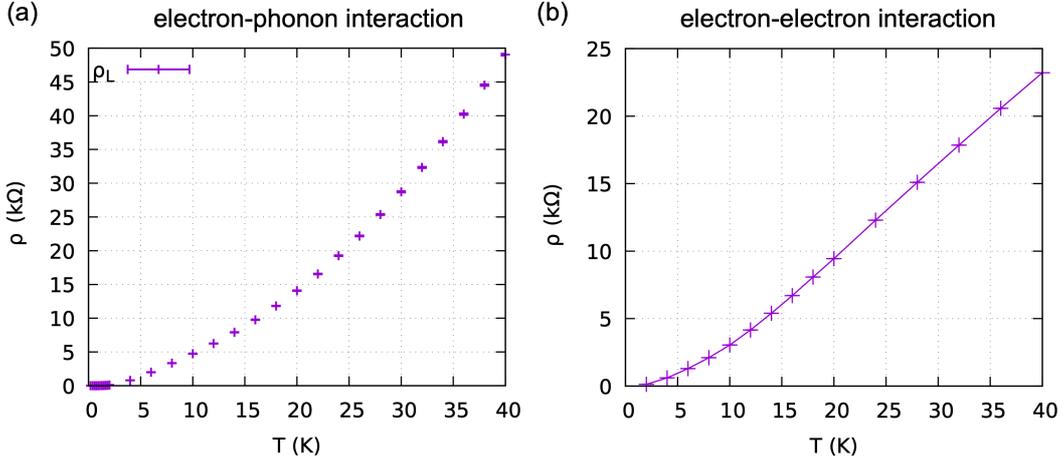

**Supplementary Figure S 8.** Temperature dependence of the resistivity for MATBG at $\theta = 1.12°$. (a) The phonon resistivity for the sound velocity $s = 1.5 \times 10^4$ ms$^{-1}$, orbital radius $\xi = a/6\sqrt{3}$ (see the form factor below Eq. 3 for details; $a$ is the lattice parameter) and $g^2 = 9.37$ eVÅ$^2$, which corresponds to the deformation potential $D = 20$ eV and mass density $1.52 \times 10^{-10}$ kg cm$^{-2}$. (b) The resistivity due to electron-electron scattering for $v/a^2 = 1.5$ meV. In both figures, we used the bandwidth $W = 3t = 3.9$ meV, the lattice parameter $a = 12.8$ nm, and the filling $\nu = 2/3$.

reads

$$H_{\mathrm{p}} = \sum_{\vec{q}} \hbar \omega_{\vec{q}} \hat{a}^\dagger_{\vec{q}} \hat{a}_{\vec{q}}, \quad \omega_{\vec{q}} = s|\vec{q}|, \qquad (2)$$

where $\hbar\omega_{\vec{q}}$ is the energy of the phonon with momentum $\vec{q}$ and the sum over $\vec{q}$ runs over the momenta $\vec{q}$ in the SL Brillouin zone. Recent study of phonons in moiré graphene finds a dispersion that, at low energies, is similar to that of the monolayer graphene [16]. Therefore, we use the sound velocity $s = 1.5 \times 10^4$ ms$^{-1}$, a typical value observed in graphene [16,17]. The electrons and phonons interact through deformation coupling,

$$H_{\mathrm{ep}} = -\sum_{\vec{q},\vec{k},n} \frac{g_{\vec{q}}}{\sqrt{V}} \sqrt{\hbar\omega_{\vec{q}}} \hat{\psi}^\dagger_{\vec{k},n} \hat{\psi}_{\vec{k},n} (\hat{a}_{\vec{q}} + \hat{a}^\dagger_{-\vec{q}}). \qquad (3)$$

Here, $V = \sqrt{3}a^2 N/2$ is the area of the system, $N$ is the number of the SL unit cells, and $g_{\vec{q}} = g\,e^{-q^2\xi^2/4}$ is the form factor where $\xi$ is a parameter reflecting the Wannier function radius; this formfactor, with $\xi = 1/5 - 1/6$, reproduces that of the Wannier function of TBG [15].

The resistivity $\rho(T)$ of a system is related to entropy production. We use the variational method assuming the conventional form of the field-induced modulation of electron distribution $\delta f_{n\vec{k}} = f_{n\vec{k}} - f^0_{n\vec{k}} = \tau e \vec{E} \cdot \vec{v}_{n\vec{k}} \beta f^0_{n\vec{k}}(1 - f^0_{n\vec{k}})$, where $f_{n\vec{k}}$ and $f^0_{n\vec{k}}$ are respectively the electron occupation rate and Fermi distribution for electrons on $n$th band and momentum $\vec{k}$. within this method, the resistivity reads [18]

$$\rho(T) = \frac{\sum_{n,n'} \int \frac{dk^2}{(2\pi)^2} \frac{dk'^2}{(2\pi)^2} \beta P^{(\text{e-ph})}_{n\vec{k},\vec{k'}-\vec{k};n'\vec{k'}} \left(v^x_{n\vec{k}} - v^x_{n'\vec{k'}}\right)^2}{\left[e \sum_n \int \frac{dk^2}{(2\pi)^2} (v^x_{n\vec{k}})^2 \beta f^0_{n\vec{k}}(1 - f^0_{n\vec{k}})\right]^2}, \qquad (4)$$

where

$$P^{(\text{e-ph})}_{n\vec{k},\vec{q};n'\vec{k'}} = 2\pi \delta_{n,n'} |g_{\vec{q}}|^2 \omega_{\vec{q}} N_{\vec{q}} f^0_{n\vec{k}}\left(1 - f^0_{n'\vec{k'}}\right) \delta(\varepsilon_{n\vec{k}} + \hbar\omega_{\vec{q}} - \varepsilon_{n'\vec{k'}}). \qquad (5)$$

Here, $\varepsilon_{\vec{k}s} = s|h_{\vec{k}}|$ and $v^x_{\vec{k}s} = \partial_{k_x} \varepsilon_{\vec{k}s}/\hbar$ are respectively the eigenenergy and the group velocity of electron state with momentum $\vec{k}$ and band index $s$, $\beta = 1/T$ is the inverse temperature, $\vec{E}$ is the external electric field, and $\tau$ is the variational parameter. Figure S8.b shows the result of resistivity by electron-phonon scattering. The linear-$T$ resistivity appears at around $T \sim 10$ K. However, the temperature dependence becomes super-linear at $T \lesssim 5$ K, as expected from the typical behavior of phonon resistivity in metals. The resistivity also deviates from a $T$-linear dependence at $T \gtrsim 30$ K due to the narrow bandwidth. The narrow window of $T$-linear resistivity cannot account for our experiment, in which the $T$-linear resistivity appears from $T \sim 20$ K down to $T = 40$ mK.

Besides the conventional acoustic-phonon contribution [14], other works point out a larger contributions from gauge phonon [19] and umklapp scattering [15]. The gauge-phonon contribution, however, shows a trend qualitatively similar to the acoustic phonons; it is $\rho \propto T^1$ in a wide range of temperature around $T \sim T_{BG}$, whereas they are nonlinear in the low temperature $T/T_{BG} \ll 1$. The umklapp scattering contribution become dominant at temperatures $T \gtrsim T_{BG}$ while their contribution is negligible in the low temperature. Other studies associate the non-Fermi liquid behavior of TBG to purely electronic phenomena [20,21]. Hence, we conclude that the electron-phonon scat-tering cannot account for the $T$-linear resistivity down to 20 mK, even considering the contribution of unconventional scattering mechanisms.

### B. Resistivity due to electron-electron scattering

We next turn to the resistivity by electron-electron scattering. In metals, the contribution from electron-electron interaction is often very small due to small umklapp scattering and their temperature dependence is $\rho \propto T^2$. To confirm the above behavior, we performed the calculation of resistivity using the variational method used for resistivity by electron-phonon interaction. The resistivity reads

$$\rho_{\text{int}}(T) = \frac{1}{4k_B T e^2} \frac{\sum_{s_i} \int \prod_i \frac{dk_i^2}{(2\pi)^2} Q^{(\text{int})}_{\vec{k}_1 s_1, \vec{k}_3 s_3; \vec{k}_2 s_2, \vec{k}_4 s_4} \left[ v^x_{\vec{k}_1 s_1} + v^x_{\vec{k}_3 s_3} - v^x_{\vec{k}_2 s_2} - v^x_{\vec{k}_4 s_4} \right]^2}{\left\{ \int \frac{dk^2}{(2\pi)^2} (v^x_{\vec{k}s})^2 \frac{\partial f^0_{\vec{k}s}}{\partial \varepsilon_{\vec{k}s}} \right\}^2}. \tag{6}$$

$$Q^{(\text{int})}_{\vec{k}_1 s_1, \vec{k}_3 s_3; \vec{k}_2 s_2, \vec{k}_4 s_4} = \frac{(2\pi)^3}{\hbar} \left( \frac{U_{\vec{k}_1 - \vec{k}_2}}{4} \right)^2 \frac{1 + s_1 s_2 s_3 s_4 \cos[s_1 \phi_{\vec{k}_1} - s_2 \phi_{\vec{k}_2} + s_3 \phi_{\vec{k}_3} - s_4 \phi_{\vec{k}_4}]}{2}$$
$$\times f^0_{\vec{k}_1 s_1} f^0_{\vec{k}_3 s_3} (1 - f^0_{\vec{k}_2 s_2})(1 - f^0_{\vec{k}_4 s_4}) \delta(\vec{k}_1 + \vec{k}_3 - \vec{k}_2 - \vec{k}_4) \delta(\varepsilon_{\vec{k}_1 s_1} + \varepsilon_{\vec{k}_3 s_3} - \varepsilon_{\vec{k}_2 s_2} - \varepsilon_{\vec{k}_4 s_4}). \tag{7}$$

Here, $\vec{G}$ is the reciprocal lattice vectors, $\hbar$ is the Dirac constant, and $U_{\vec{q}}$ is the Fourier transform of electron-electron interaction. For simplicity, we approximate the momentum dependence of scattering rate by $v^2 = (U_{\vec{q}}/4)^2$. Note that this resistivity is the usual contribution purely coming from electron-electron interaction, and not those from the non-Mathiessen contribution; the latter contribution appears in the presence of both impurity and electron-electron scatterings giving $T^2$ temperature dependence at a very low temperature. Figure S8.a shows the temperature dependence of the resistivity for electron-electron scattering, which shows a nonlinear curve below $T \lesssim 20$ K. The nonlinear temperature dependence is consistent with the common notion that the temperature dependence is $\rho \propto T^2$ when electron-electron scattering dominates. However, the calculated result is qualitatively different from what observed in our experiment, excluding the possibility of conventional electron-electron scattering mechanism.

# SI References

[1] Y. Cao et al. Strange metal in magic-angle graphene with near Planckian dissipation. Physical Review Letters 124, 076801 (2020).
[2] H. Polshyn et al. Large linear-in-temperature resistivity in twisted bilayer graphene. Nature Physics 15, 1011-1016 (2019).
[3] K. Kadowaki &. S. B. Woods. Universal relationship of the resistivity and specific heat in heavy-Fermion compounds. Solid State Communications 58, 507 (1986).
[4] X. Lin, B. Fauqué & K. Behnia. Scalable $T^2$ resistivity in a small single-component Fermi surface. Science 349, 945-948 (2015).
[5] J. R. Wallbank et al. Excess resistivity in graphene superlattices caused by umklapp electron-electron scattering. Nature Physics 15, 32-36 (2019).
[6] S. A. Hartnoll & A. P. Mackenzie. Planckian dissipation in metals. Preprint arXiv:2107.07802 (2021).
[7] U. Zondiner et al. Cascade of phase transitions and Dirac revivals in magic-angle graphene. Nature 582, 203-208 (2020).
[8] J. Custers et al. The break-up of heavy electrons at a quantum critical point. Nature 424, 524-527 (2003).
[9] N. Nagaosa & P. A. Lee. Normal-state properties of the uniform resonating-valence-bond state. Physical Review Letters 64, 2450 (1990).
[10] C. Collignon et al. Heavy non degenerate electrons in doped strontium titanate. Physical Review X 10, 031025 (2020).
[11] I. M. Hayes et al. Scaling between magnetic field and temperature in the high-temperature superconductor $BaFe_2(As_{1-x}P_x)_2$. Nature Physics 12, 916-919 (2016).
[12] P. Giraldo-Gallo, J. A. Galvis, et al. Scale-invariant magnetoresistancein a cuprate superconductor. Science 361, 479-481 (2019).
[13] P.A. Lee. Low-temperature T-linear resistivity due to umklapp scattering from a critical mode. Physical Review B 104, 035140 (2021).
[14] F. Wu, E. Hwang & S. D. Sarma. Phonon-induced giant linear-in-T resistivity in magic angle twisted bilayer graphene: Ordinary strangeness and exotic superconductivity. Physical Review B 99, 165112 (2019).
[15] H. Ishizuka et al. Purcell-like enhancement of electron-phonon interactions in long-period superlattices: linear-temperature resistivity and cooling power. Nano Letters, 21, 18, 7465-7471 (2021).
[16] M. Koshino and N. N. T. Nam. Effective continuum model for relaxed twisted bilayer graphene and moiré electron-phonon interaction. Physical Review B 101, 195425 (2019).
[17] X. Cong et al., Carbon 149, 19-24 (2019).
[18] J. M. Ziman, Electrons and Phonons Reprint ed. (Oxford Univ. Press, Oxford) (2001).
[19] I. Yudhistira et al., Gauge-phonon dominated resistivity in twisted bilayer graphene near magic angle. Physical Review B 99, 140302 (2019).
[20] J. Gonzàlez & T. Stauber. Marginal Fermi liquid in twisted bilayer graphene. Physical Review Letters 124, 186801 (2020).
[21] P. Cha, A. A. Patel & E. A. Kim. Strange metals from melting correlated insulators in twisted bilayer graphene. Physical Review Letters 127, 266601 (2021).
[22] J. A. N. Bruin, H. Sakai, R. S. Perry & A. P. Mackenzie. Similarity of scattering rates in metals showing T-linear resistivity. Science 339 (6121), 804-807 (2013).